\documentclass[preprint,superscriptaddress, nofootinbib]{revtex4}%
\usepackage{graphicx}
\usepackage{color}
\usepackage[tight]{units}

\definecolor{light-gray}{rgb}{0.8,0.8,0.8}

\begin{document}

\title{Electron-phonon coupling in single-layer MoS$_2$}
\author{Sanjoy K. Mahatha}
\email{sanjoymahatha@gmail.com}
\affiliation{Department of Physics and Astronomy, Interdisciplinary Nanoscience Center (iNANO), Aarhus University, 8000 Aarhus C, Denmark}
\affiliation{Deutsches Elektronen-Synchrotron DESY, Hamburg, Germany}
\author{Arlette S. Ngankeu}
\affiliation{Department of Physics and Astronomy, Interdisciplinary Nanoscience Center (iNANO), Aarhus University, 8000 Aarhus C, Denmark}
\author{Nicki Frank Hinsche}
\affiliation{Center for Atomic-scale Materials Design, Department of Physics, Technical University of Denmark, DK-2800 Kgs. Lyngby, Denmark}
\affiliation{Department of Physics, Martin-Luther-University Halle-Wittenberg, 06099 Halle, Germany}
\author{Ingrid Mertig}
\affiliation{Department of Physics, Martin-Luther-University Halle-Wittenberg, 06099 Halle, Germany}
\affiliation{Max Planck Institute of Microstructure Physics, 06120 Halle, Germany}
\author{Kevin Guilloy}
\author{Peter L. Matzen}
\author{Marco Bianchi}
\affiliation{Department of Physics and Astronomy, Interdisciplinary Nanoscience Center (iNANO), Aarhus University, 8000 Aarhus C, Denmark}
\author{Charlotte E. Sanders}
\affiliation{Central Laser Facility STFC Rutherford Appleton Laboratory, Harwell, Didcot OX11 0QX, United Kingdom}
\author{Jill A. Miwa}
\affiliation{Department of Physics and Astronomy, Interdisciplinary Nanoscience Center (iNANO), Aarhus University, 8000 Aarhus C, Denmark}
\author{Harsh Bana}
\affiliation{Physics Department, University of Trieste, Via Valerio 2, 34127 Trieste, Italy}
\author{Elisabetta Travaglia}
\affiliation{Physics Department, University of Trieste, Via Valerio 2, 34127 Trieste, Italy}
\author{Paolo Lacovig}
\affiliation{Elettra - Sincrotrone Trieste S.C.p.A., AREA Science Park, S.S. 14 km 163.5, 34149 Trieste, Italy}
\author{Luca Bignardi}
\affiliation{Elettra - Sincrotrone Trieste S.C.p.A., AREA Science Park, S.S. 14 km 163.5, 34149 Trieste, Italy}
\author{Daniel Lizzit}
\affiliation{Elettra - Sincrotrone Trieste S.C.p.A., AREA Science Park, S.S. 14 km 163.5, 34149 Trieste, Italy}
\author{Rosanna Larciprete}
\affiliation{CNR-Institute for Complex Systems, Via Fosso del Cavaliere 100, 00133 Roma, Italy}
\author{Alessandro Baraldi}
\affiliation{Physics Department, University of Trieste, Via Valerio 2, 34127 Trieste, Italy}
\affiliation{Elettra - Sincrotrone Trieste S.C.p.A., AREA Science Park, S.S. 14 km 163.5, 34149 Trieste, Italy}
\affiliation{IOM-CNR, Laboratorio TASC, AREA Science Park, S.S. 14 km 163.5, 34149 Trieste, Italy}
\author{Silvano Lizzit}
\affiliation{Elettra - Sincrotrone Trieste S.C.p.A., AREA Science Park, S.S. 14 km 163.5, 34149 Trieste, Italy}
\author{Philip Hofmann}
\email{philip@phys.au.dk}
\affiliation{Department of Physics and Astronomy, Interdisciplinary Nanoscience Center (iNANO), Aarhus University, 8000 Aarhus C, Denmark}

\begin{abstract}
The electron-phonon coupling strength in the spin-split valence band maximum of single-layer MoS$_2$ is studied using angle-resolved photoemission spectroscopy and density functional theory-based calculations. Values of the electron-phonon coupling parameter $\lambda$ are obtained by measuring the linewidth of the spin-split bands as a function of temperature and fitting the data points using a Debye model. The experimental values of $\lambda$ for the upper and lower spin-split bands at K are found to be 0.05 and 0.32, respectively, in excellent agreement with the calculated values for a free-standing single-layer MoS$_2$. The results are discussed in the context of spin and phase-space restricted scattering channels, as reported earlier for single-layer WS$_2$ on Au(111). The fact that the absolute valence band maximum in single-layer MoS$_2$ at K is almost degenerate with the local valence band maximum at $\Gamma$ can potentially be used to tune the strength of the electron-phonon interaction in this material.
\end{abstract}

\maketitle
\section{Introduction}

Two-dimensional semiconducting materials, in particular single-layer (SL) transition metal dichalcogenoides (TMDCs) have attracted tremendous interest due to their resemblance to graphene, but possessing a sizeable band gap and unique optical properties \cite{Bollinger:2001aa,Mak:2010aa,Splendiani:2010aa}. For example, SL MoS$_2$ \cite{Mak:2010aa, Splendiani:2010aa,Cao:2012ab,Radisavljevic:2011aa,Bruix:2016aa}, which is one of the most studied TMDCs, has a direct band gap, in contrast to its bulk counterpart, and correspondingly different optical properties \cite{Mak:2010aa, Splendiani:2010aa}. Moreover, the lack of inversion symmetry in its trigonal prismatic structure lifts the spin degeneracy at the K and K$^{\prime}$ valleys and gives the opportunity to exploit coupled spin and valley degrees of freedom \cite{Mak:2014aa,Xu:2014ac,Schaibley:2016}.

High quality SL TMDCs can be grown on different substrates instead of isolation by micro-mechanical exfoliation. It has recently become possible to grow large area \cite{Miwa:2015aa, Sorensen:2014aa}, single orientation epitaxial SL MoS$_2$  \cite{Bana:2018} and WS$_2$ \cite{Bignardi:2018aa} on Au(111). The very high quality of these samples offers the opportunity to study the spin texture \cite{Eickholt:2018aa} and electron-phonon coupling strength \cite{Hinsche:2017} near the valence band (VB) maximum at the K and K$^{\prime}$ points experimentally and thus allowing access to parameters relevant for transport in hole-doped devices \cite{Hsu:2015aa} or low-dimensional superconductivity \cite{Costanzo:2016aa}. Of particular interest is the spin-splitting near the VB top, as this entails the possibility to have a different electron-phonon coupling strengths for two states that mainly differ by their spin polarization. 

Recently, a dramatically different electron-phonon coupling strength was demonstrated for the strongly spin-split VB of WS$_2$ on Au(111) \cite{Hinsche:2017} and explained in terms of phase space restriction and spin-selective scattering. In particular, a very weak electron-phonon coupling was found for the ``upper'' spin-split VB (the one forming the absolute VB maximum). This can be understood by the large energy separation of the VB maximum from other states, implying that the holes in the VB maximum at K cannot decay by electron-phonon coupling because either the available states for scattering are well outside a phonon energy window or the transitions (to the equivalent band at K$^{\prime}$) are spin forbidden.

In this paper, we present experimental and theoretical results on the electron-phonon coupling strengths on the spin-split VB of SL MoS$_2$ at K and along the K-$\Gamma$ direction. SL MoS$_2$ is structurally and electronically quite similar to WS$_2$ but with two significant differences: The first is the much smaller spin-orbit splitting of the VB near K and the second is the fact that that the local VB maximum at $\Gamma$  is almost degenerate with the absolute VB maximum at K. These differences can affect the available scattering channels for electron-phonon coupling and could thus lead to appreciable differences between MoS$_2$ and WS$_2$. We find experimentally that the electron-phonon coupling for the VB maximum at K is still very weak and that the experimental values for the coupling strength in the spin-split branches  are in excellent agreement with the  ones calculated for free standing SL MoS$_2$. However, the weak coupling is predicted to hold only for a small region of k-space around the K point. We show that this is related to the position of the local VB maximum at $\Gamma$ and we discuss how the position of this band can be changed in order to tune the electron-phonon coupling in a wider window around K. 

\section{Methods}

\subsection{Sample Preparation and Characterization}

Single-layer MoS$_2$ was synthesized at the SuperESCA beamline at Elettra \cite{Baraldi:2003ab} using a well-established procedure of Mo evaporation onto the clean Au(111) surface in a background pressure of H$_2$S \cite{Lauritsen:2007aa,Miwa:2015aa,Dendzik:2015aa}. Very high crystalline quality and a single orientation of the MoS$_2$ layer with respect to the substrate was achieved by a careful optimisation of the growth parameters \cite{Bana:2018}. Angle-resolved photoemission (ARPES) spectra were acquired at the SGM-3 beamline of ASTRID2 \cite{Hoffmann:2004aa} using a photon energy of 30~eV. The total energy and angular resolutions were better than 30~meV and 0.2$^{\circ}$, respectively.

\subsection{Theoretical Methods}
The electronic structure of single-layer free standing MoS$_2$ was calculated using density functional theory as implemented in a modified \textsc{Quantum\-Espresso} suite \cite{Giannozzi:2017}. The vibrational properties were obtained within density functional perturbation theory using the same package. 
Relativistic effects, \emph{i. e.} spin-orbit coupling, were treated self-consistently and are accounted for in all calculations. The optimized norm-conserving pseudo-potentials were generated with the \textsc{ONCVPSP} \cite{Hamann:2013bq} code and a kinetic energy cutoff for the wavefunction expansion of 68 Ry was used for all calculations\footnote{Sulphure atomic configuration: [Ne] 3s$^2$ 3p$^4$ 3d$^{-2}$, Molybdenum atomic configuration: [Kr] 4d$^5$ 5s$^1$ 5p$^{0}$}. Exchange and correlation effects were accurately accounted for by the PBE flavour of the general gradient approximation \cite{Perdew:1996}, while long-range dispersion corrections were semi-empirically considered \cite{Grimme:2006} to specifically account for a proper description of the inter-layer atomic relaxations.

For the self-consistent computations Monkhorst-Pack meshes for the electronic Brillouin zone (BZ) integration ($k$) and for the phononic BZ integration ($q$) have been set to $18 \times18 \times 1~k$ as well as $18 \times18 \times 1~k$ and $9 \times 9 \times 1~q$, for the electron and (electron-)phonon calculations, respectively. The self-consistent calculations were reiterated until the root mean square change of the total energy became smaller than $10^{-10}$ Hartree.\footnote{Note that for a proper description of a non-imaginary ZA phonon mode a rather small threshold for self-consistency of $10^{-15}$ was needed at an electronic smearing of  $0.01$Ry.}.

The optimized lattice constant for freestanding SL MoS$_2$ was found to be $a_{\text{MoS$_2$}}=3.19$~{\AA} with a vertical spacing  of $d_{\text{S-S}}=1.57$~{\AA} between the two sulphur layers in the material. This corresponds to an enlargement of the in-plane lattice constant of almost 1\% compared to the bulk value, with no noticeable change of $d_{\text{S-S}}$. 
The atomic relaxation can have a significant effect on the energy difference of the VB maximum at $\Gamma$ with respect to the upper VB maximum at K, as sketched in Figure~\ref{model}. Values of a few up to almost 100~meV have been stated in previous calculations and a possible impact on the electron-phonon coupling will be discussed later \cite{Zhu:2011ad,Kormanyos:2013aa,Rasmussen:2015aa}.
 
The electron-phonon induced broadening of the electronic states was obtained within a modified version of the \textsc{EPW} code \cite{Ponce:2016}. An improved tetrahedron Fermi-surface-adaptive integration scheme based on the Wannier-interpolated electron-phonon matrix elements was applied \cite{Kawamura:2014,Assmann:2016,Hinsche:2018}. Wannier interpolated electronic bands had an error of less than $5$ meV compared to the \textit{ab initio} derived eigenvalues.
The \textit{a priori} state-dependent, temperature-dependent phonon-induced electronic linewidth $\Gamma_{n\mathbf{k}}$ is closely related to the imaginary part of the lowest order electron-phonon self-energy $\Sigma_{n\mathbf{k}}''$ for the Bloch state of energy 
$\varepsilon_{n\mathbf{k}}$ at band $n$ and momentum $\mathbf{k}$ and given by 
\begin{eqnarray}\label{equ:ellw}
\Gamma_{n\mathbf{k}}(T) =  2\Sigma_{n\mathbf{k}}''(T) = 2\pi \sum_{m\nu} \int_{\rm BZ} \frac{d\mathbf{q}}{\Omega_{\rm BZ}} | g_{mn,\nu}(\mathbf{k,q}) |^2 \nonumber \\
    \times   \Big\{\big[n_{\mathbf{q}\nu}(T)+f_{m\mathbf{k+q}}(T)\big]\delta(\varepsilon_{n\mathbf{k}}-\varepsilon_{m\mathbf{k+q}} +\omega_{\mathbf{q}\nu}) \nonumber \\
+ \big[n_{\mathbf{q}\nu}(T)+1-f_{m\mathbf{k+q}}(T)\big]\delta(\varepsilon_{n\mathbf{k}} -\varepsilon_{m\mathbf{k+q}} - \omega_{\mathbf{q}\nu})\Big\},
\end{eqnarray}
where $g_{mn,\nu}(\mathbf{k,q})$ is the electron-phonon scattering matrix element; $n$ and $f$ are Bose and Fermi functions, respectively; $\varepsilon$ and $\omega$ are non-interacting electron and phonon energies, respectively. 
The electron-phonon coupling parameter $ \lambda_{n\mathbf{k}}$ is essentially counting the possible scattering processes for a chosen initial state $(\mathbf{k},n)$ into possible final states, weighted by a squared matrix element:
\begin{eqnarray}\label{equ:lambda}
   \lambda_{n\mathbf{k}} = \sum_{m\nu} \int_{\rm BZ} \frac{d\mathbf{q}}{\Omega_{\rm BZ} \omega_{\mathbf{q}\nu}} | g_{mn,\nu}(\mathbf{k,q}) |^2 \nonumber \\
    \times  \delta(\varepsilon_{n\mathbf{k}}-\varepsilon_{m\mathbf{k+q}} \pm \omega_{\mathbf{q}\nu}).
\end{eqnarray}  
For a chosen initial state $(\mathbf{k},n)$ a sum over at least 250,000 final states $m$, participating phonon modes $(\mathbf{k+q}\nu)$, and their corresponding Fourier-interpolated $\omega_{\mathbf{q}\nu}$ and Wannier-interpolated electron-phonon matrix elements $g_{mn,\nu}(\mathbf{k,q})$ was performed to assure convergence. 
 
\section{Results and discussion}
\begin{figure}[h]
\includegraphics[width=0.6\textwidth]{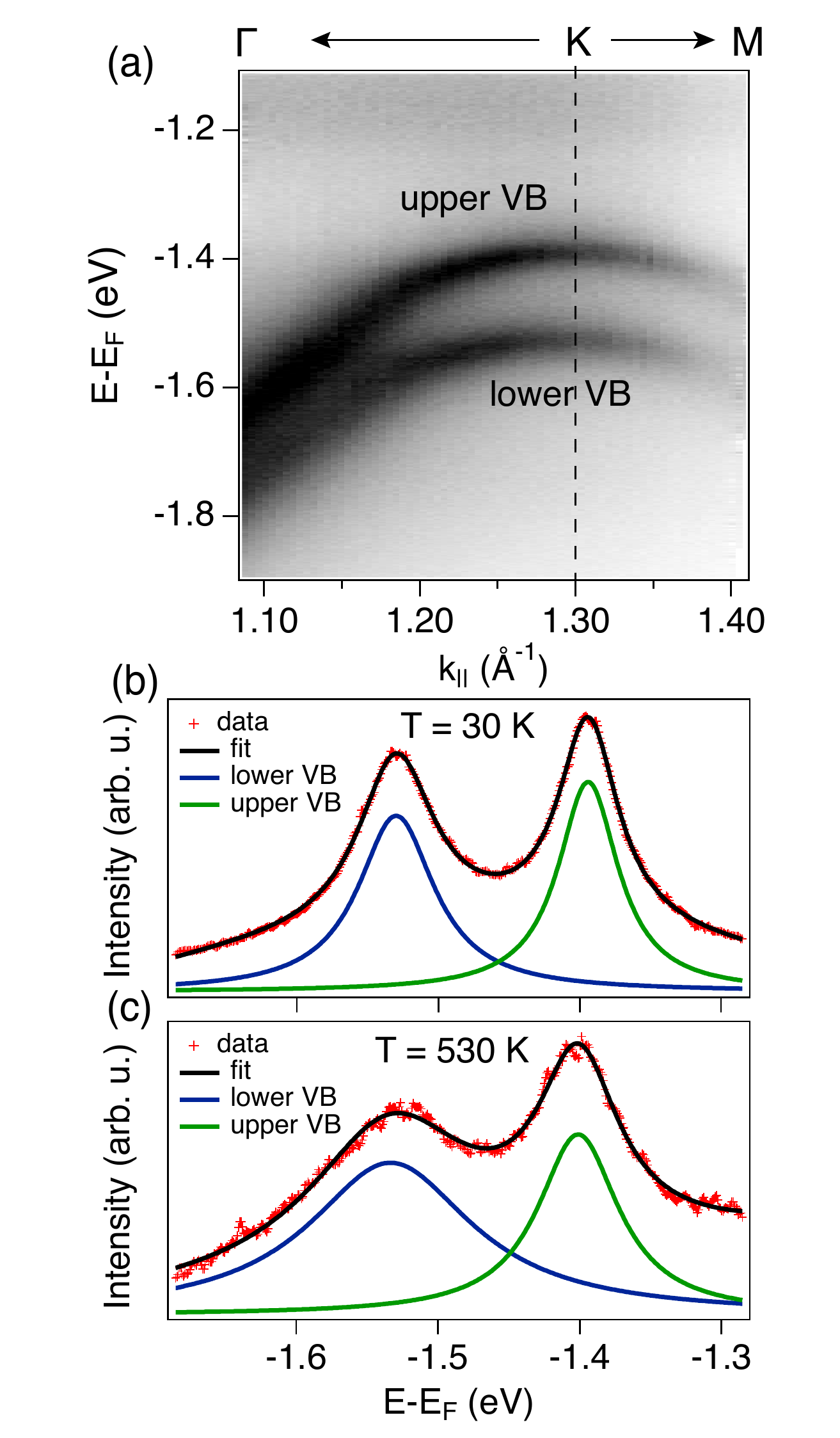}\\
\caption{(a) Photoemission intensity of SL MoS2/Au(111) along the $\Gamma$-K-M  direction acquired at 30~K. (b, c) Energy distribution curves  through the  K point at low (30~K) and high (530~K) temperature, respectively, with a fit using two Lorentzian peaks and a polynomial background.}
\label{ARPES}
\end{figure}

Figure \ref{ARPES}(a) shows the photoemission intensity of SL MoS$_2$ grown on Au(111) near the K point, showing the two well-separated spin-split branches of the VB which are labelled as upper VB and lower VB. It should be noted that both  bands are very narrow, indicating a high degree of structural order. Moreover, the two bands lie in the projected band gap of Au(111) \cite{Takeuchi:1991aa}, and one can thus rule out the direct hybridization with the substrate. 

In ARPES, the electron-phonon coupling strength for any state in the band structure can be extracted by measuring the temperature-dependent linewith at the desired $k$-point \cite{Hofmann:2009ab} and this procedure is applied here for the upper and lower VB maximum, similar to the previously reported results for SL WS$_2$ on Au(111)\cite{Hinsche:2017}. Figure \ref{ARPES}(b) and (c) show a comparison of two energy distribution curves (EDCs) through the K point, taken at 30~K and 530~K, together with a fit to the data using two Lorentzian peaks with a polynomial background  in order to extract the temperature-dependent linewidths $\Gamma_{exp}$.  The much higher temperature in Fig. \ref{ARPES}(c)  leads to an electron-phonon scattering-induced broadening of the two peaks but the broadening is clearly much more pronounced for the lower VB. 

\begin{figure}[h]
\includegraphics[width=0.8\textwidth]{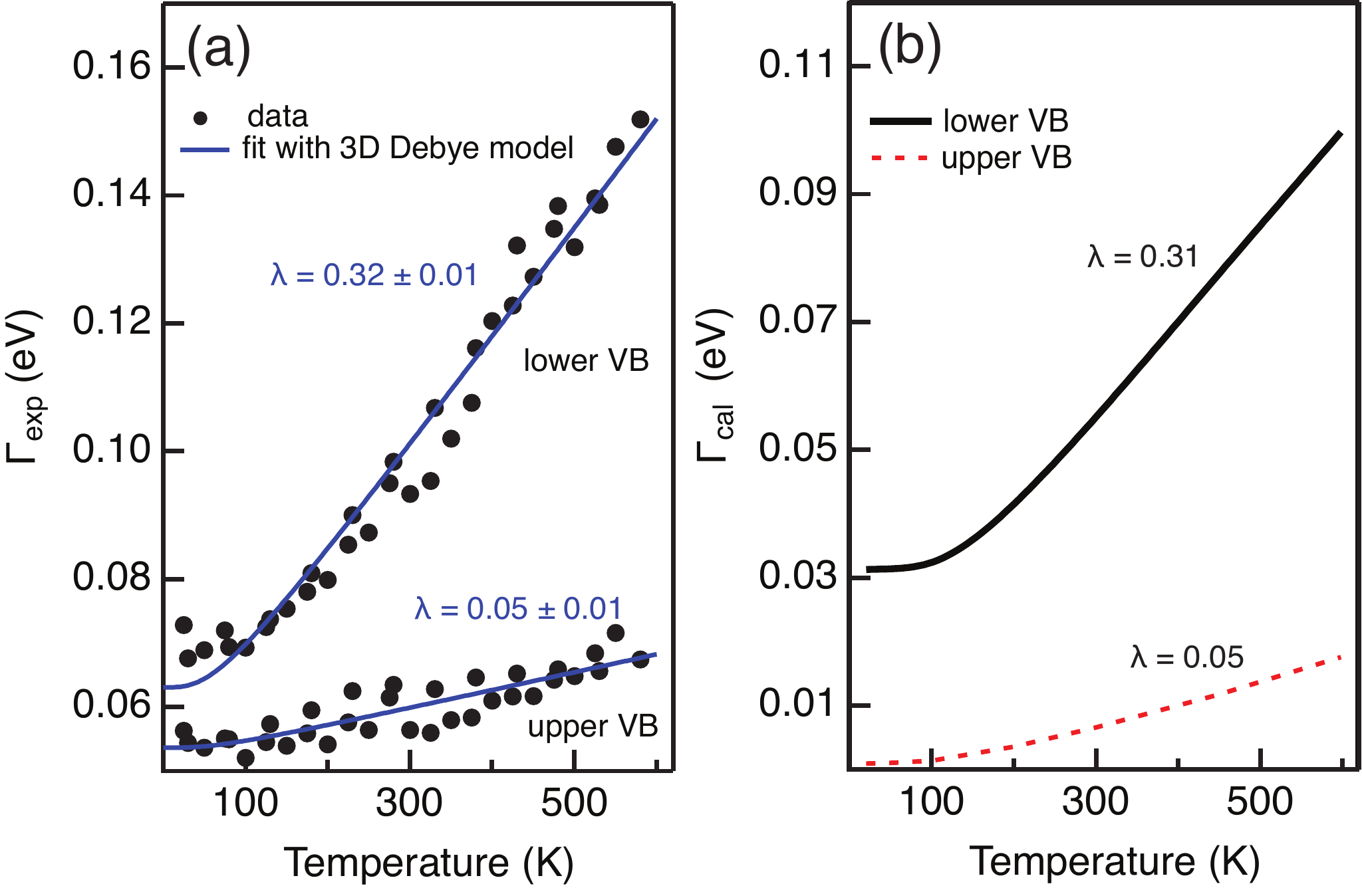}\\
\caption{(a) Temperature-dependent Lorentzian linewidth of the upper and lower VB at K (filled black circles). The electron-phonon contributions to the linewidth calculated using 3D Debye model are shown by the blue lines. A constant offset has been added to allow for the contributions from electron-electron and electron-defect scattering. (b) Calculated electron-phonon scattering contribution to the temperature-dependent linewidths of the spin-split VB branches at K for free-standing SL MoS$_2$.}
\label{ElPh_exp}
\end{figure}

The temperature-dependence of  $\Gamma_{exp}$  for upper and lower VB is given by the black filled circles in Figure \ref{ElPh_exp}(a). Note that to acquire the data for this study, a total of two ascending and one descending temperature series was performed. The solid lines show a fit to the data using a 3D Debye model with a Debye temperature of 262.3~K \cite{Peng:2016}. The fit includes a temperature-independent offset in order to account for the effects of electron-electron and electron-defect scattering. The resulting values for the electron-phonon coupling strength $\lambda$ are 0.05(1) and 0.32(1) for the upper and lower band, respectively. Note that the choice of the model (Debye or Einstein) and  its dimensionality (2D or 3D) does not have a major effect of the resulting $\lambda$ values because $\lambda$ is essentially equivalent to the slope of the curves above the Debye temperature for all models \cite{Hofmann:2009ab}.  Note also that the quoted uncertainties refer only to the particular fit and do not represent the uncertainties that could arise from the choice of model or Debye temperature.

Figure \ref{ElPh_exp}(b) shows the corresponding calculated temperature-dependent linewidths for the spin-split upper VB and lower VB at K for free-standing SL MoS$_2$. The calculation does not include effects of electron-electron and electron-defect scattering and it is therefore not meaningful to compare the absolute linewidth values to the experiment. However, since the electron-phonon is the only linewidth contribution with a significant temperature dependence, the temperature-induced changes in the linewidth can be compared. In particular, the electron-phonon coupling strength  $\lambda$ can be extracted  from the high-temperature part of the  calculated curves, via a linear fit above $T$=450~K. The $\lambda$ values of 0.05 and 0.31 obtained for the upper and lower VB, respectively, fit extremely well with the experimental results.

\begin{figure}[h]
\includegraphics[width=0.8\textwidth]{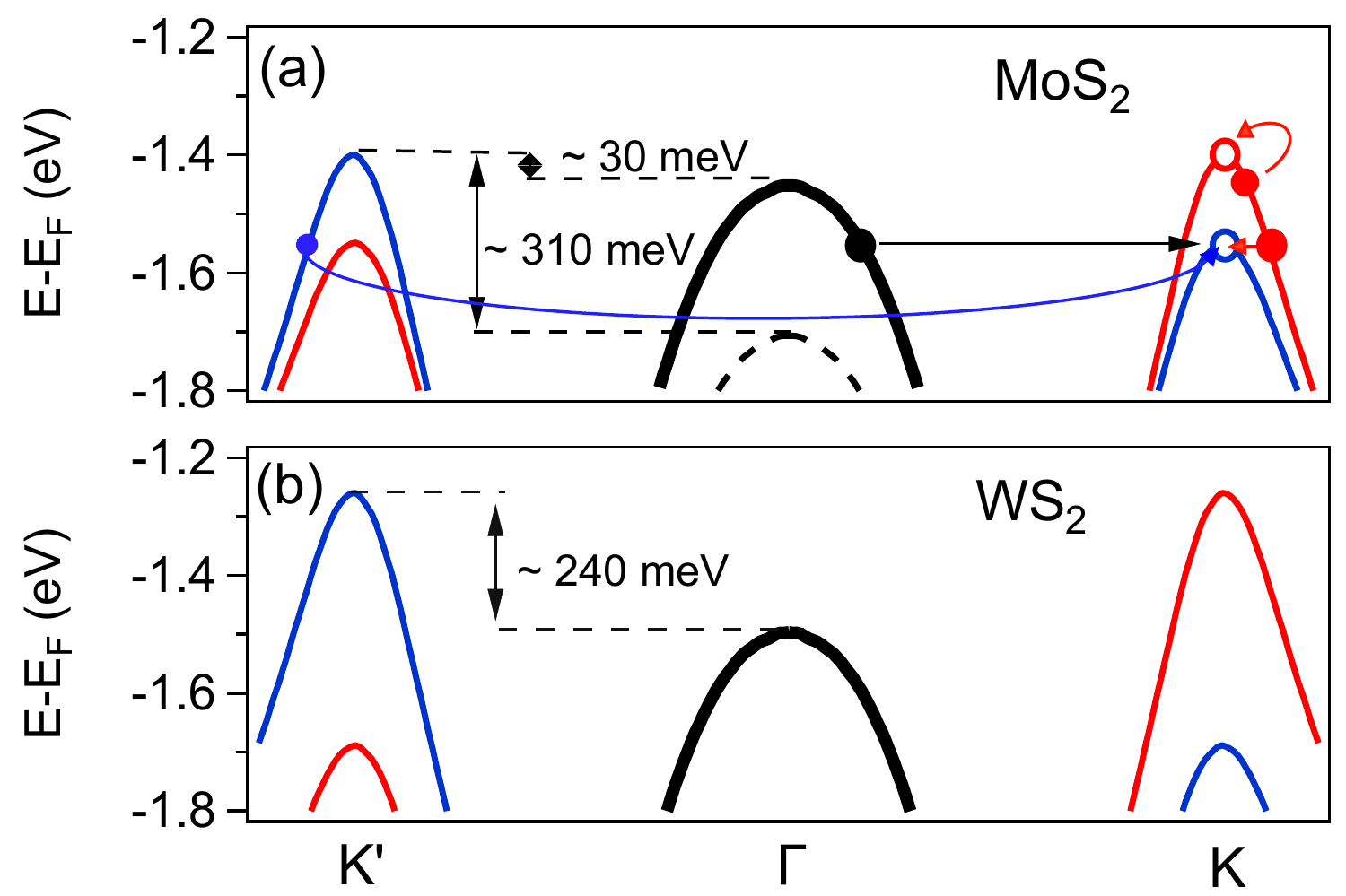}\\
\caption{Sketch of the VB for (a) SL MoS$_2$ and (b) SL WS$_2$ with a colour coding indicating the spin directions at K and K$^{\prime}$, and a spin degenerate local maximum at $\Gamma$. A hole in the upper VB at K can only be filled using an electron at higher binding energy from the same band, and from states near $\Gamma$ if the phonon energy exceeds 30 meV. Contrary, a hole near the top of the lower VB can be filled involving electrons from the VB near $\Gamma$ or from the upper branch of the VB at K$^{\prime}$. Other possible phonon emission and absorption processes are omitted for clarity. The black dashed band in (a) indicates the approximate position of the VB at $\Gamma$ for MoS$_2$ on Au(111).}
 \label{model}
\end{figure}

The pronounced difference in the electron-phonon coupling strength for the two VB branches is comparable to what has been observed on SL WS$_2$ \cite{Hinsche:2017} and can be qualitatively understood by the sketch of the band structure of SL MoS$_2$ in Figure \ref{model}(a). 
Consider a hole at the maximum of the upper VB at K. Using a phonon to provide energy and momentum, the hole can only be filled by an electron from the same band at higher binding energy by annihilating a thermally excited phonon. The hole cannot be filled using an electron from the corresponding band at K$^{\prime}$ because such a transition would be spin-forbidden, or from the lower spin-split band at K because these states are well outside the phonon energy window. It can be filled by an electron from the band at $\Gamma$ if the phonon energy exceeds 30~meV. However the latter contributions are small and obviously unavailable at low temperatures. This greatly restricts the phase space for electron-phonon scattering and also explains why the calculated linewidth goes to zero for low temperatures: In the absence of thermally excited phonons the lifetime of the hole diverges. 

The situation is entirely different for a hole in the lower VB at K. This can be filled by processes involving electrons from either the VB at $\Gamma$, or from the upper band at K$^{\prime}$, where the spin-polarization is retained. Scattering to the upper spin-split VB at K is also possible (the spin-polarization decreases for states not exactly at K and K$^{\prime}$), but it is very inefficient as the scattering process involves an almost full spin-flip. A key-difference to the upper VB is that the transitions can involve an electron at lower binding energy and phonon excitation rather than annihilation. The result is a much stronger electron-phonon coupling and a finite linewidth at zero temperature. 


\begin{figure}[h]
\includegraphics[width=0.6\textwidth]{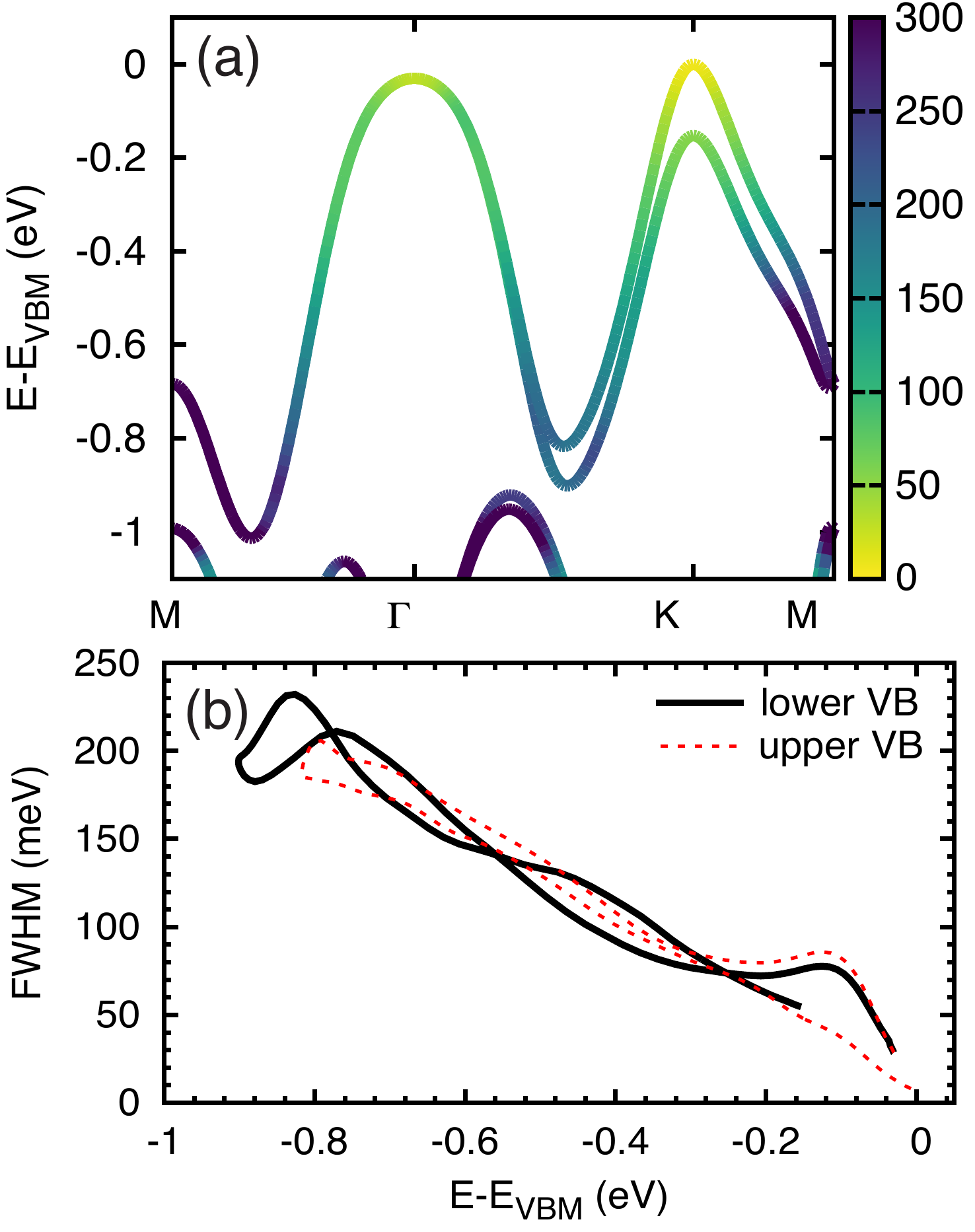}\\
\caption{(a) Band structure with colour-scale encoded linewidths of SL MoS$_2$ VB at a temperature of 300 K. (b) Quantitative representation of the same data for the spin-split upper and lower VB along $\Gamma$-K.}
  \label{ElPh_cal}
\end{figure} 

A complete picture of the electron-phonon coupling of SL MoS$_2$ is given in Figure \ref{ElPh_cal}(a), showing the electron-phonon scattering-induced linewidth at 300~K for the upper VB along high-symmetry lines in the Brillouin zone. The linewidth is encoded in the colour of the band structure plot. A quantitative plot of the data along $\Gamma$K is given in Figure \ref{ElPh_cal}(b). The weak coupling strength for the upper band is clearly confined to a small $k$ and energy region around K. As soon as the energy drops below that of the band maximum at $\Gamma$, the linewidth increases significantly. Interestingly, coupling at the band maximum near $\Gamma$ is also markedly weaker than at higher binding energies, suggesting that scattering within the band must play an important role, similar to the situation in graphene \cite{Mazzola:2013aa,Mazzola:2017aa}. 

\begin{figure}[h]
\includegraphics[width=0.8\textwidth]{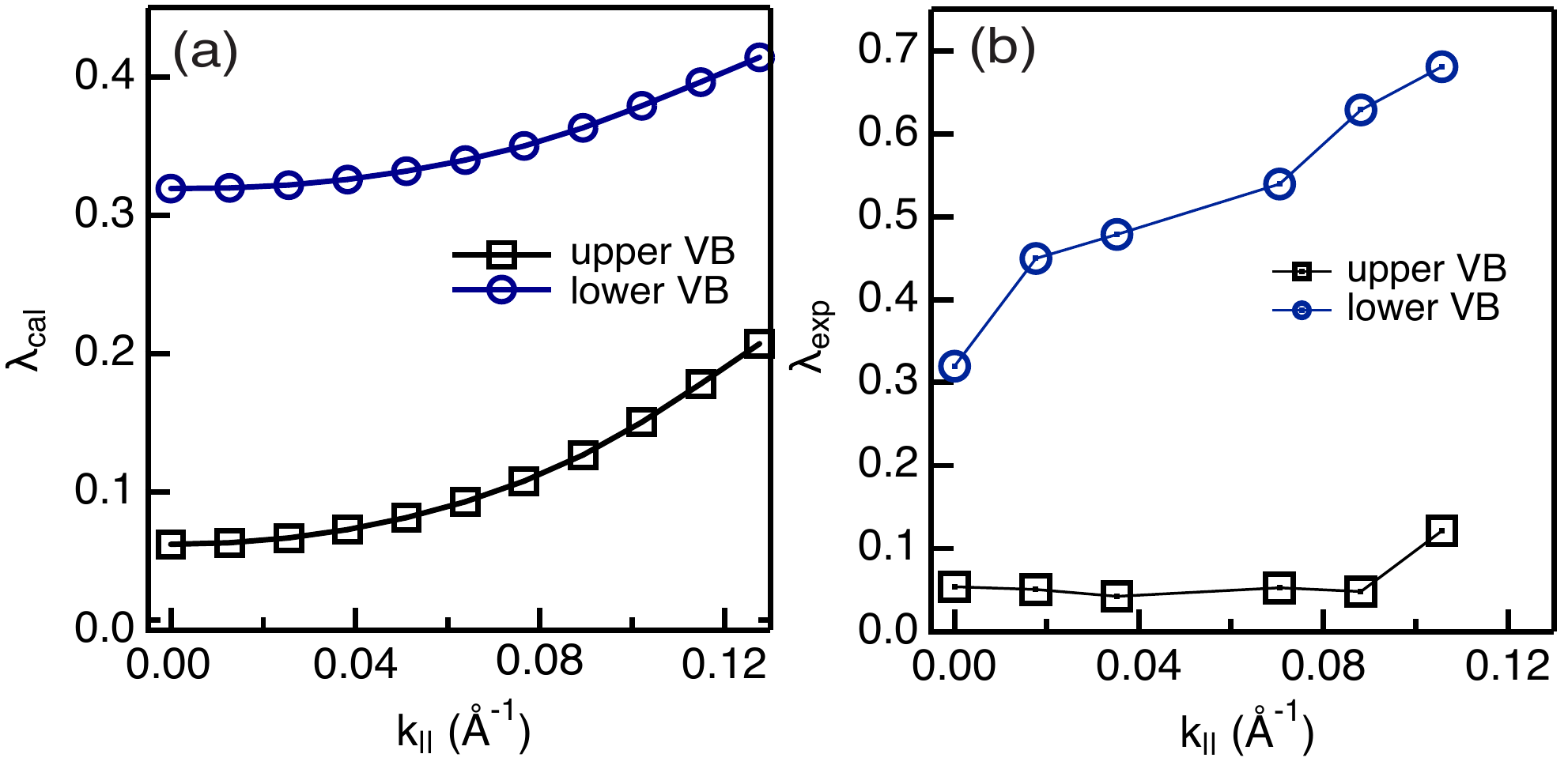}\\
\caption{(a) Calculated and (b) experimental electron-phonon coupling constant $\lambda$ as a function of crystal momentum along the K-$\Gamma$ direction. Note that the K point here has been set to zero. The experimental values of $\lambda$ have been determined by a fit to a Debye model corresponding to that in Fig. \ref{ElPh_exp}(a).}
 \label{lambda_kg}
\end{figure}

When comparing these results to the case of SL WS$_2$  \cite{Hinsche:2017}, we find some interesting similarities and differences arising from the different band alignment in the two systems. A sketch of the band structure for SL WS$_2$ is given in Fig. \ref{model}(b). The most important difference to SL MoS$_2$ is the greatly increased spin-splitting due to the heavier W atoms. This leads to a very large separation of the bands at K of 419~meV \cite{Dendzik:2015aa} such that electron-phonon scattering from the upper into the lower band is forbidden for a wide $k$ range around K. The separation from the band at $\Gamma$ is also large and the combination of these two effects results in a very weak electron-phonon coupling for the upper VB in a wide range around K.  For SL MoS$_2$ we also find this weak coupling, albeit in a smaller $k$-range. Here, the VB at $\Gamma$ is much closer to the the upper band at K and allows scattering by phonons with energies above 30~meV. This delicate detail of the band alignment allows for an additional scattering channel for states in the upper VB and thus explains a larger electron-phonon coupling compared to SL WS$_2$. There are several factors that can influence this band alignment towards both stronger and weaker overlap. The first is the choice of substrate that determines the degree of hybridization and the direction of the hybridization-induced band shifts \cite{Bruix:2016aa}. This includes the option to reduce  the substrate - TMDC interaction by intercalation \cite{Mahatha:2018aa}. The second option is the application of biaxial mechanical strain that can be used to seamlessly tune the energy difference of the valence bands at K and $\Gamma$ within the given stability of the material \cite{Lloyd:2016aa, Peelaers:2012aa,Moghadasi:2017aa}. Compressive strain shifts the band at $\Gamma$ downwards while tensile strain shifts it upwards. Shifts in the order of several hundred meV can be easily realized \cite{Moghadasi:2017aa}. Finally, the band alignment can be influenced by external fields \cite{Shanavas:2015hh,Cheng:2016gz,Zhao:2017ct,Affandi:2018cc}.

For the particular case of MoS$_2$ on Au(111), the interaction with the substrate moves the (hybridized) VB at $\Gamma$ to a significantly higher binding energy, such that the energy difference between the band maxima at K and $\Gamma$ is about 310~meV \cite{Miwa:2015aa,Bruix:2016aa} (for a sketch of this situation, see the dashed band at $\Gamma$ in Fig. \ref{model}(a)). For such a relative band shift, the electron-phonon coupling between the upper VB at K and the band at $\Gamma$ should only start to matter at a higher binding energy and one would thus expect an increased $k$-range of weak electron-phonon coupling around K. Figure \ref{lambda_kg} shows a direct test of this idea, based on comparing the calculated electron-phonon coupling constant for free standing SL MoS$_2$ (with the $\Gamma$ band corresponding to the solid line in Fig. \ref{model}(a)) and the experimentally determined $\lambda$ values for SL MoS$_2$ on Au(111) (with the $\Gamma$ band corresponding to the dashed line in Fig. \ref{model}(a)). An inspection of $\lambda$ for the upper band appears to confirm the expected trend: While the agreement between experiment and calculations is excellent near K, the calculated $\lambda$ quickly increases when moving away from K, whereas the experimentally determined $\lambda$ remains very low for a wider range of $k$. For the lower band, on the other hand, $\lambda$ increases much more rapidly in the experimental data than expected for the free-standing layer. The origin of this is not clear but it could be related to an increased coupling to the broad spectrum of hybridized Au-MoS$_2$ states near $\Gamma$ \cite{Bruix:2016aa}. 

\section{Conclusions}
In summary, we have investigated the electron-phonon coupling strength in the spin-split VB near K in SL MoS$_2$ using experimental data from SL MoS$_2$ on Au(111) and calculations for a free-standing layer. Calculation and experiment show excellent agreement at the K point and the findings can be explained  using simple phase space arguments. The electron-phonon coupling is very weak for the upper VB at K, i.e. for the states that are mainly responsible for hole transport in the material. This weak coupling results from reduced phase space for scattering, similar to the situation for SL WS$_2$. However, the energy difference between the band maxima at K and $\Gamma$ responsible for preventing the inter-band scattering is small for SL MoS$_2$ and can be tuned by several parameters. This potentially permits to tailor the electron-phonon interaction strength in order to either favour strong coupling (for superconductivity) or weak coupling (for transport in conventional devices). We have demonstrated this by showing that a weak coupling for the upper VB can be found for a wider $k$ range around K for SL MoS$_2$ on Au(111) than expected for the free standing layer, consistent with the change of band alignment due to hybridization. 

\section{acknowledgement}

This work was supported by the Danish Council for Independent Research, Natural Sciences under the Sapere Aude program (Grants No. DFF-4002-00029 and DFF-6108-00409) and by VILLUM FONDEN via the Centre of Excellence for Dirac Materials (Grant No. 11744) and the Aarhus University Research Foundation.  Affiliation with the Center for Integrated Materials Research (iMAT) at Aarhus University is gratefully acknowledged. NFH received funding within the H.C. \O rsted Programme from the European Union's Seventh Framework Programme and Horizon 2020 Research and Innovation Programme under Marie Sklodowska-Curie Actions grant no. 609405 (FP7) and 713683 (H2020). The Center for Nanostructured Graphene (CNG) is sponsored by the Danish National Research Foundation, Project No. DNRF103.

\end{document}